\documentstyle[12pt,fleqn]{article}
\def\beq{\begin{equation}}
\def\eeq{\end{equation}}
\def\half{\mbox{$1\over2$}}
\def\0{\otimes}
\def\1{\lambda}
\def\2{_\lambda}
\def\6{\langle}
\def\9{\rangle}
\def\Du{D_{uv}}
\def\Dx{D_{xy}}

\begin{document}

\begin{center}

{\large{\bf Optimal eavesdropping in quantum cryptography. I.}}\\[7mm]

Christopher A. Fuchs,$^1$ Nicolas Gisin,$^2$ Robert B.
Griffiths,$^3$\\ Chi-Sheng Niu,$^3$ and Asher Peres$^{4,*}$\\[7mm]
{\it $^1$Norman Bridge Laboratory of Physics 12-33, California
Institute of Technology, Pasadena, CA 91125\\
$^2$Group of Applied Physics, University of Geneva, CH 1211 Geneva 4,
Switzerland\\
$^3$Department of Physics, Carnegie-Mellon University, Pittsburgh, PA
15213\\
$^4$Institute for Theoretical Physics, University of California,
Santa Barbara, CA 93106}\\[7mm]
{\bf Abstract}\bigskip\end{center}

We consider the Bennett-Brassard cryptographic scheme, which uses two
conjugate quantum bases. An eavesdropper who attempts to obtain
information on qubits sent in one of the bases causes a disturbance
to qubits sent in the other basis. We derive an upper bound to the
accessible information in one basis, for a given error rate in the
conjugate basis. Independently fixing the error rate in the conjugate
bases, we show that both bounds can be attained simultaneously by an
optimal eavesdropping probe, consisting of two qubits. The qubits' 
interaction and their subsequent measurement are described
explicitly. These results are combined to give an expression for the
optimal information an eavesdropper can obtain for a given average
disturbance when her interaction and measurements are performed
signal by signal. Finally, the relation between quantum cryptography
and violations of Bell's inequalities is discussed.\vfill

\noindent PACS number(s): 03.65.-w, 42.79.Sz, 89.70.+c\vfill

\noindent $^*$Permanent address: Department of Physics,
Technion---Israel Institute of Technology, 32\,000 Haifa,
Israel\newpage

\begin{center}{\bf I. INTRODUCTION}\end{center}\medskip

In quantum cryptography, individual quanta are prepared in
nonorthogonal quantum states to encode and carry information about
cryptographic keys.  In this way, an eavesdropper can acquire
information about the key only at the risk of causing a detectable 
disturbance. The oldest and best known cryptographic scheme, BB84, is
due to Bennett and Brassard \cite{BB84}: the information sender,
called Alice, encodes each logical bit, 0 or 1, into the linear
polarization of a single photon, along one of two conjugate bases of
her choice, as shown in Fig.~1. The receiver, Bob, measures the
polarization of the photon in one of the two bases, either $xy$ or
$uv$, randomly chosen by him. Only after that, Alice reveals to him
the basis she used.  This information is sent on a public channel
that can be monitored, but not modified, by anyone else. Bob then
likewise tells Alice whether he used the correct basis. If he did,
Alice and Bob know one bit, that no one else ought to know.

After this protocol has been repeated many times, Alice and Bob
sacrifice some of these secret bits by publicly comparing their
values.  This gives them an estimate of the noise on the channel,
which may be due to either natural causes or to the presence of an
eavesdropper (Eve).  In the latter case, the maximal amount of
information that Eve could have gathered is, in principle, fixed by
the laws of quantum mechanics.  If Eve's information is small enough 
compared to the noise she has induced, Alice and Bob may still be
able to use {\it classical\/} methods of privacy
amplification~\cite{BBBSS,EHPP} in order to reduce Eve's information
to an arbitrarily small level. It is therefore important
to estimate the maximal amount of information that Eve may have
acquired, for a given error rate observed by Bob.

There are many possible strategies for eavesdropping, some of which
have been analyzed by other authors. Ekert and Huttner~\cite{EH}
examined a simple ``intercept-resend'' method, where Eve performs
standard von~Neumann measurements. L\"utkenhaus~\cite{Lu} considered
the use of positive operator-valued measures (POVM)~\cite{qt} under
the restriction that Eve performs her measurements before Alice
reveals the basis. Recently, Gisin and Huttner~\cite{GH} determined
the optimal strategy for an eavesdropper restricted to a
two-dimensional probe (a single qubit) interacting on line with each
transmitted signal, with the probe measured after the basis is
revealed. These various results, and the optimal ones obtained
in the present article, are plotted in Fig.~2.

The common denominator of all these strategies is that they are 
restricted to interactions and measurements on each individual signal
sent from Alice to Bob; there are no ``collective'' interactions or 
measurements on strings of signals, as might be the case if Eve were 
able perform quantum measurements on systems of arbitrary size. 
Furthermore, none of the strategies allow Eve to delay her
measurements until the completion of Alice and Bob's privacy
amplification, and none take into account the information
leaked to her during the public communication phase of the protocol.
The latter kind of information depends upon which bits are ultimately
discarded and upon the specific algorithm used in the privacy 
amplification process.  Finally, even within the restrictions set
by this paradigm, none of the schemes can claim optimality in the
sense of specifying the best possible ratio between Eve's information
gain and her induced disturbance.

The purpose of this article is twofold.  The first is to give a 
quantitative statement of the physical principle responsible for the 
operation of the BB84 protocol: an eavesdropper who attempts to
obtain information in one basis causes a disturbance to the conjugate
basis.  The second---more relevant to practical quantum
cryptography---is to derive the absolute best achievable information
an eavesdropper can obtain about a single qubit, for a given average
error rate caused to the signals.  In both these tasks, we again work
within the paradigm cited above.  Namely, we assume that Eve may
interact with only one 
signal at a time and may only make measurements on each individual 
probe.  Furthermore, she may do this after Alice announces her basis,
but before the execution of any error testing or privacy
amplification protocols.

{}From the point of view of ultimate security in cryptography, these 
restrictions may be severe.  On the other hand, with respect to 
experimental science, these assumptions are hardly limiting at all.
Indeed it is only now becoming possible to make two qubits interact
with one another in a controlled fashion \cite{expwork}; controllable
interactions between three qubits, as would be required for the
optimal strategy presented here, are still quite some way in the
future. Finally, though an expression for the tradeoff between
information and disturbance in a less restrictive scenario may be
eminently important for cryptography, such a relation---because of
its dependence on the details of privacy amplification---cannot be
fundamental and lies somewhat beyond the scope of basic physics.

The plan of our paper is as follows.
In Sect.~2, we derive a general bound that refers to the
accessible information in one basis, corresponding to a given error
rate in the {\it conjugate\/} basis.  This is obtained without using
any particular model for the eavesdropping interaction; the latter is
assumed only to be unitary.  It had been previously known that a
four-dimensional probe (that is, one consisting of two interacting 
qubits) is the largest needed for achieving the optimal detection of 
signals emitted in a two-dimensional space~\cite{FP}. There are,
however, some cases for which a two-dimensional probe is
sufficient~\cite{FP}. It turns out that, for the BB84 protocol, the
bound can only be attained by eavesdropping probes consisting of two
qubits. The optimal interaction, and the subsequent measurement
protocol are described explicitly in Sect.~3.  In particular, it is
shown that, upon independently fixing the error rates in each basis,
there is an optimal eavesdropping strategy for achieving the two
bounds simultaneously.  A quantum computational circuit representing
the optimal strategy is described in the following paper~\cite{GN}.

Finally, in Sect.~4, we address issues directly relevant to quantum 
cryptography by constructing the optimal tradeoff relation for Eve's
overall accessible information in terms of the average error rate for
both bases.  This is obtained by two methods.  The first relies on
the work of the previous two sections; the second incorporates an
argument based on a symmetrization technique.
Note that both Sect.~2 and 3 concern fundamental physical questions.
The ``practically minded'' cryptographer need only browse through
them, and may then proceed directly to Sect.~4 to find results
relevant to privacy amplification~\cite{BBBSS}. In the concluding
remarks we return to fundamental physics by outlining an intriguing
connection between the optimal information--disturbance tradeoff and
a violation of Bell's inequality in the Bennett-Brassard-Mermin
modification of the BB84 protocol~\cite{BBM}. This confirms an idea
first expressed by Ekert~\cite{Ekert91} and recently made
quantitative by Gisin and Huttner~\cite{GH}.

\bigskip\begin{center}{\bf II. INFORMATION AND DISTURBANCE IN
CONJUGATE BASES}\end{center}\medskip

If Eve performs standard (von Neumann type) measurements in the
$xy$ basis, she does not disturb signals sent in that basis, but
she disturbs maximally those sent in the $uv$ basis, and
vice-versa. In this section, it will be shown that, quite generally,
Eve's ability to obtain partial information on the signals sent in
one of the bases is related to the disturbance caused to the signals
sent in the other basis.  This is relevant to eavesdropping on the
BB84 protocol because it is the raw physical fact that allows its
operation.

We take the framework for our problem directly from quantum 
cryptography. In order to take advantage of Alice's delayed
information on the basis that was used, Eve's optimal strategy is the
following: she lets a probe, initially in some standard state
$|\psi_0\9$, interact unitarily with the qubit sent by Alice. (There
is no loss of generality in this, because any physical nonunitary
interaction is equivalent to a unitary one with a higher dimensional
probe). Eve's probe is then stored until Alice announces the basis
that was used, and only after that is it measured by Eve.

In a convenient notation, if Alice sends state $|x\9$, the
result may be written as

\beq |x\9\0|\psi_0\9\to|X\9, \eeq
where $|X\9$ is an entangled state of the probe and the photon that
Alice sent to Bob.  Likewise, for the other signals that Alice may
send, the results of Eve's intervention are entangled states,
$|Y\9$, $|U\9$, and $|V\9$. Since the interaction is unitary, 
it follows from

\beq |x\9=(|u\9+|v\9)/\sqrt{2}\qquad{\rm and}\qquad |y\9=
(|u\9-|v\9)/\sqrt{2}, \label{conj} \eeq
that

\beq |X\9=(|U\9+|V\9)/\sqrt{2}\qquad{\rm and}\qquad |Y\9=
(|U\9-|V\9)/\sqrt{2}.  \label{CONJ} \eeq
Eve's measurement on the probe may be of the standard type (an
orthogonal projection valued measure) or, more generally, it may be
of the POVM type~\cite{qt}, where the various outcomes correspond to
a set of positive semi-definite operators that sum to the identity 
operator on the probe's Hilbert space.  Since Eve waits until Alice
reveals her basis, she may choose a POVM $\{E_\1\}$ when the $xy$
basis is sent, and a different POVM $\{F_\1\}$ when the $uv$  basis
is sent.

Note that the interaction of Eve's probe with the qubit sent by Alice
to Bob completely determines the mean error rate for signals sent in
the $xy$ basis and those in the $uv$ basis.  It also determines Eve's
accessible information (i.e., her maximal information) for both types
of signals.  The aim of this section is to show that the accessible
information for $xy$ signals is simply related to the mean error rate
for $uv$ signals, and vice-versa.  These mean values are well
defined regardless of which signal is sent in any single instance.
In particular, there is nothing counterfactual about comparing the
information in one basis with the error rate in the other basis.

Let us now set about our task.  If Alice sent a signal 
$|x\9$, the probability that Eve detects outcome $\1$ is

\beq P_{\1x}=\6X|{\bf1}\0E\2|X\9, \label{Plx}\eeq
and likewise for the other signals. Here, {\bf1} is the
identity operator for Alice and Bob's qubit. Let $p_i$ be the
prior probability that Alice sends signal $i$. The probability
that Eve gets outcome $\1$ when Alice uses the $xy$ basis is thus

\beq q\2=P_{\1x}\,p_x+P_{\1y}\,p_y. \label{q}\eeq
If Eve observes outcome $\1$ when she tests her probe, the posterior
probability (or {\it likelihood\/}) Eve assigns signal $i$ is, by
Bayes' theorem,

\beq Q_{i\1}=P_{\1i}\,p_i/q\2. \label{post}\eeq

How can Eve make use of this result? One possibility is to simply
assume that the largest one of $Q_{x\1}$ and $Q_{y\1}$ indicates
the signal that was actually sent by Alice. Then, the smallest one
of $Q_{x\1}$ and $Q_{y\1}$ is Eve's expected error rate. A convenient
measure of her information gain is~\cite{Fuchs96}

\beq G\2=|Q_{x\1}-Q_{y\1}|. \label{Gl} \eeq
For example, this expression would be Eve's expected income, if she
were earning one dollar for each correct guess, and paying one dollar
for each incorrect guess. This expression is also related in a simple
way to Eve's expected error rate~\cite{Fuchs96} in her interpretation
of the result $\1$, which is $\half\,(1-G\2)$.

On average, Eve's information gain (in bits) is

\beq \sum\2 q\2\,G\2=\sum\2|P_{\1x}\,p_x-P_{\1y}\,p_y|.\eeq
If the two signals are equiprobable, Eve's average gain is

\beq G=\half\,\sum\2|P_{\1x}-P_{\1y}|, \label{G}\eeq
and her expected average error rate is $\half\,(1-G)$.

A more sophisticated data processing by Eve is to keep track of all
the $q\2$ and $Q_{i\1}$ of her observations. These may then
be used to compute her {\it mutual information\/} on Alice's
message~\cite{qt}.  With equiprobable signals, this is given
(in nats) by

\beq I=\ln2+\sum\2 q\2 \sum_i Q_{i\1}\,\ln Q_{i\1}. \label{I} \eeq
This measure of Eve's information is the main concern of this
article. However, in the following, 
we shall consider first the simple ``information gain'' expression
(\ref{G}), for which a bound is easier to find. This result
will then be used to bound the mutual information.

Let us first consider the case where Alice announced
that she had sent a signal in the $xy$ basis, and Eve observed
outcome $\1$. We then have, from Eqs.~(\ref{post}) and (\ref{Gl}),

\beq q\2\,G\2=\half\,|P_{\1x}-P_{\1y}|
 =\half\,|\6X|{\bf1}\0 E\2|X\9-\6Y|{\bf1}\0 E\2|Y\9|. 
\eeq
This can also be written, thanks to Eq.~(\ref{CONJ}), as
\begin{eqnarray} q\2\,G\2 & = & \half\,|\6U|{\bf1}\0 E\2|\,V\9+
  \6V|{\bf1}\0 E\2|\,U\9|,\nonumber \\
 & = & |{\rm Re} \6U| B_u \0 E\2|\,V\9
  + {\rm Re} \6U| B_v \0 E\2|\,V\9|, \nonumber \\
 & \leq & |\6 U_{\1u} | V_{\1u}\9| + | \6U_{\1v} |
V_{\1v}\9|
 \label{real}\end{eqnarray}
where $B_u = |u\9 \6u|$ and $B_v = |v\9 \6 v|$ are projectors onto
Bob's states  $|u\9$  and $|v\9$, so that
\beq
 B_u+B_v={\bf 1},
\eeq
 and  
\begin{eqnarray}
 | U_{\1u}\9 & = & B_u \0 \sqrt{E\2}\, |U\9,\qquad 
 | V_{\1u}\9 = B_u \0 \sqrt{E\2}\, |V\9,\nonumber \\
 | U_{\1v}\9 & = & B_v \0 \sqrt{E\2}\, |U\9,\qquad 
 | V_{\1v}\9 = B_v \0 \sqrt{E\2}\, |V\9.
\label{UVdefs}
\end{eqnarray}
Note that $\sqrt{E\2}$ is well defined, since $E\2$ is a positive
semi-definite operator. Of course, $\sqrt{E\2}$ can be replaced by $E\2$
when $E\2$ is a projector.  

The Schwarz inequality implies that

\beq
 |\6 U_{\1u} | V_{\1u}\9| \leq [\6 U_{\1u} | U_{\1u}\9
\6 V_{\1u} | V_{\1u}\9]^{1/2} \label{Schwarz}
\eeq
with equality if and only if $ | U_{\1u}\9$ and
$| V_{\1u}\9$ are parallel. The physical meaning of the
expression $\6 V_{\1u} | V_{\1u}\9$  is that, if instead of
the scenario considered here,
Alice had actually sent signal $|v\9$, Eve would get result $\1$ and
Bob would get $|u\9$ (that is, a wrong result) with a probability
equal to that expression. Therefore, we shall write

\beq
\6 V_{\1u}| V_{\1u}\9 = P_{\1v}\,d_{\1v}\qquad{\rm and}
 \qquad \6 V_{\1v} |V_{\1v}\9 = P_{\1v}(1-d_{\1v}),
\eeq
where $P_{\1v}$ is
defined as in Eq.~(\ref{Plx}), and $d_{\1v}$ is the probability
that Bob gets a wrong result {\it conditioned upon\/} Alice sending
$|v\9$ and Eve measuring $\1$. The other terms in Eq.~(\ref{real})
can be handled in the same way, and we finally obtain

\beq q\2\,G\2\le\sqrt{P_{\1u}\,P_{\1v}}\,
 \left[\sqrt{d_{\1v}\,(1-d_{\1u})}+
 \sqrt{d_{\1u}\,(1-d_{\1v})}\,\right].
\label{Gl2}
\eeq

Let us develop the bound in Eq.~(\ref{Gl2}) further.  By the
geometric mean -- arithmetic mean inequality, we have

\beq (P_{\1u}\,P_{\1v})^{1/2}\leq\half\,(P_{\1u}+P_{\1v})
 =q\2,
\label{Gmean}
\eeq
where the first equality holds if $P_{\1u}=P_{\1v}$, and where
Eq.~(\ref{q}) was used.  Let us now define $d\2$ and $w$ by

\beq 
d_{\1u}=d\2+w\qquad{\rm and}\qquad d_{\1v}=d\2-w.
\label{Difd}
\eeq
The square bracket in Eq.~(\ref{Gl2}) is easily seen to be an even
function of $w$, which has its maximum value at $w=0$, that is, when
$d_{\1u}=d_{\1v}=d\2$.  That is to say, the bound reaches a
maximum when the probability of detectable disturbance is identical
for each of the conjugate basis vectors. We thus have

\beq G\2\leq2\,[d\2\,(1-d\2)]^{1/2}. \label{Gd}\eeq
It follows that Eve's information gain averaged over all
outcomes is bounded by the expression

\beq G=\sum\2 q\2\,G\2\leq2\sum\2 q\2\,[d\2\,(1-d\2)]^{1/2}.\eeq
Since the function $[x(1-x)]^{1/2}$ is concave, we have~\cite{ineq}

\beq
\sum\2 q\2\,[d\2\,(1-d\2)]^{1/2}\leq[D\,(1-D)]^{1/2},
\label{Jensen}
\eeq
where $D=\sum q\2 d\2$ is Bob's observable error rate, i.e., the one
averaged over all of Eve's outcomes. Equality holds
only if all the $d\2$ are equal to $D$. Thus, finally,

\beq
G_{xy}\leq2\,[D_{uv}\,(1-D_{uv})]^{1/2},
\label{Mildred}
\eeq
where the indices have been introduced to emphasize that Eve's
information gain refers to signals sent in the $xy$ basis, and
Bob's error rate refers to signals sent in the $uv$ basis.

In exactly the same fashion as above, we can derive a bound on the
information gain with respect to the the $xy$ basis in terms of
the disturbance inflicted upon the $uv$ basis:

\beq
G_{uv}\leq2\,[D_{xy}\,(1-D_{xy})]^{1/2}.
\label{Leon}
\eeq

Equations~(\ref{Mildred}) and (\ref{Leon}) tell us that Eve's maximal
information gain, for given error rate caused to Bob in the conjugate
basis, is bounded in a simple way.  The main goal of this section,
however, is in finding an analogous bound on the mutual information 
$I$, defined by Eq.~(\ref{I}). The latter can be expressed more
simply by writing

\beq Q_{x\1}=(1+r\2)/2\qquad{\rm and}\qquad Q_{y\1}=(1-r\2)/2,\eeq
since these two expressions sum to unity. We then have

\beq I=\half\,\sum\2 q\2\,[(1+r\2)\ln(1+r\2)+(1-r\2)\ln(1-r\2)].
 \label{I1} \eeq
Note that

\beq r\2=Q_{x\1}-Q_{y\1}=\pm G\2,\eeq
by virtue of Eq.~(\ref{Gl}). We can therefore write, instead of
Eq.~(\ref{I1}),

\beq I=\half\,\sum\2 q\2\,[(1+G\2)\ln(1+G\2)+(1-G\2)\ln(1-G\2)].
 \label{I2} \eeq

To obtain a bound on $I$, it is convenient to define a function

\beq \phi(z)=(1+z)\ln(1+z)+(1-z)\ln(1-z).\eeq
Since $\phi'(z)=\ln[(1+z)/(1-z)]$ is positive for $0<z<1$, we see 
that the right hand side of Eq.~(\ref{I2}) will increase if we
replace $G\2$ by a larger expression, such as the right hand side
of Eq.~(\ref{Gd}).  Therefore,

\beq
I\leq\half\,\sum\2 q\2\,\phi\Bigl[2\sqrt{d\2\,(1-d\2)}\,\Bigr].
\eeq
In Appendix A, it is shown that $\phi\Bigl[2\sqrt{x(1-x)}\,\Bigr]$
is a concave function of $x$.  It follows, just as in
Eq.~(\ref{Jensen}), that

\beq
I_{xy}\leq\half\,\phi\Bigl[
2\sqrt{D_{uv}\,(1-D_{uv})}\,\Bigr],
\label{ID}
\eeq
where subscripts have been added, as in Eq.~(\ref{Mildred}), to
emphasize that the information gain and error rate refer to signals
sent in two different bases.  Likewise

\beq
I_{uv}\leq\half\,\phi\Bigl[
2\sqrt{D_{xy}\,(1-D_{xy})}\,\Bigr],
\label{IDb}
\eeq
is the counterpart of Eq.~(\ref{Leon}).

Necessary and sufficient conditions for Eqs.~(\ref{ID}) and 
(\ref{IDb}) to hold as equalities are derived easily by tracing back
through the chain of inequalities that brought them about.  Let us 
focus on Eq.~(\ref{ID}).  To begin with, the concavity of 
$\phi\Bigl[2\sqrt{x(1-x)}\,\Bigr]$ is 
strict, so all the $d\2$'s must be equal; thus, in view of the 
remark following Eq.~(\ref{Difd}), we have

\beq
d_{\1u}= d_{\1v}=d\2 = D_{uv}.
\label{Melville}
\eeq
Similarly, Eq.~(\ref{Gmean}) can be a strict equality only if

\beq
P_{\1u} = P_{\1v} = q\2.
\eeq
Equality in Eq.~(\ref{real}) means that both
$\6 U_{\1u} | V_{\1u}\9$ and $\6 U_{\1v} | V_{\1v}\9$
are real and have the same sign
\begin{eqnarray}
\sigma\2 &=& 
\mbox{sign}\Bigl(\6 U_{\1u} | V_{\1u}\9 +
\6 U_{\1v} | V_{\1v}\9\Bigr),
\nonumber\\
&=&
\mbox{sign}\,(P_{\1x} - P_{\1y})\, =\,
   \mbox{sign}\,(Q_{x\1} - Q_{y\1})\, ,
\label{Sigma}
\end{eqnarray}
Finally, equality in Eq.~(\ref{Schwarz}), and its analog with $u$
replaced by $v$, means that $|U_{\1u}\9$ is a multiple of
$|V_{\1u}\9$, and $|U_{\1v}\9$ is a multiple of
$|V_{\1v}\9$.  Thus

\beq
\6V_{\1u}|V_{\1u}\9=\mu^2\6U_{\1u}|U_{\1u}\9,
\eeq
and

\beq
\6U_{\1v}|U_{\1v}\9=\nu^2\6V_{\1v}|V_{\1v}\9\;,
\eeq
for some real numbers $\mu$ and $\nu$. 

Combining these results gives the necessary and sufficient conditions
for equality in Eq.~(\ref{ID}): for every $\1$,

\beq
|V_{\1u}\9 =\epsilon\2\, \sqrt{D_{uv} \over 1 - D_{uv}}\, |U_{\1u}\9
\label{Beebub}\eeq
and

\beq
|U_{\1v}\9= \epsilon\2\, \sqrt{D_{uv} \over 1 - D_{uv}}\, |V_{\1v}\9,
\label{Conduv}
\eeq
where $\epsilon\2 = \pm 1$.

The corresponding conditions for equality in Eq.~(\ref{IDb}) are
derived in an analogous way.  Namely, if Eve uses a POVM  $\{F_\1\}$
for gaining information about the $uv$ basis---which is different
from the POVM $\{E\2\}$ used for the $xy$ basis---then the
conditions that must be satisfied are:

\beq
|Y_{\1x}\9 = \gamma\2\,\sqrt{D_{xy} \over 1 - D_{xy}}\, |X_{\1x}\9 
\label{HipHop}
\eeq
and

\beq
|X_{\1y}\9 = \gamma\2\,\sqrt{D_{xy} \over 1 - D_{xy}}\, |Y_{\1y}\9,
\label{Condxy}
\eeq
with

\beq \gamma\2\,=\,\mbox{sign}\,(P_{\1u} - P_{\1v})\,=\,
   \mbox{sign}\,(Q_{u\1} - Q_{v\1}), \label{Tau}\eeq
and
\begin{eqnarray}
 | X_{\1x}\9 & = & B_x\0 \sqrt{F\2}\, |X\9,\qquad
 | Y_{\1x}\9\, =\, B_x \0 \sqrt{F\2}\, |Y\9,\nonumber \\
 | X_{\1y}\9 & = & B_y \0 \sqrt{F\2}\, |X\9,\qquad
 | Y_{\1y}\9\, =\, B_y \0 \sqrt{F\2}\, |Y\9.
\label{XYdefs} \end{eqnarray}

In the cryptographic setting, the fact that Eve can adapt
her measurement to the basis that Alice reveals, leads one
to question whether there may be a single interaction between Eve's 
probe and Alice's qubit that saturates both Eq.~(\ref{ID}) and
Eq.~(\ref{IDb}).  We address the achievement of these bounds in the
next section.\bigskip

\begin{center}{\bf III. ATTAINABILITY OF BOTH CONJUGATE BASIS
BOUNDS}\end{center}\medskip
                                            
In this section, we show how Eve can optimize her strategy to attain
the bounds in Eqs.~(\ref{ID}) and (\ref{IDb}) with both $\Dx$ and
$\Du$ fixed independently.  The train of thought that led to the
present solution is a long and complex one.  First, we performed a
``brute force'' numerical optimization, similar to the one in
Ref.~\cite{FP}. The result was found to saturate the bound on Eve's
overall information about both bases (still to be derived in the next
section).  This led us to look for an exact analytic solution
satisfying Eqs.~(\ref{Beebub}) and (\ref{Conduv}), and (\ref{HipHop})
and (\ref{Condxy}), first with equal error rates, and then with
independent error rates. The one described below, for independent
error rates, was obtained with a certain amount of guesswork. For the
case of equal error rates, as in Section IV, there is a
symmetrization procedure that leads directly to a solution. It is
easy to check that the solution here is correct,
but the extent to which it is unique (aside from trivial changes of
basis and of phase) remains unknown.  A quantum circuit embodying the
optimal strategy is described in the following paper~\cite{GN}.

Let us fix both $\Dx$ and $\Du$.  A natural ansatz for an optimal
interaction on Eve's part is that when Alice sends a signal in the
$xy$ basis, Bob receives a simple mixture of the same two basis
vectors; when Alice sends a signal in the $uv$ basis, Bob receives a
simple mixture of these two basis  vectors. That is, Bob's density
matrix is always diagonal in the basis chosen by Alice. Then, owing
to Eq.~(\ref{Melville}) and the analogous  condition for the $xy$
basis, the Schmidt decompositions for the post-interaction states
must be of the form
\begin{eqnarray} |X\9 &=&
\sqrt{1-\Dx}|x\9|\xi_x\9+\sqrt{\Dx}|y\9|\zeta_x\9, \nonumber
\\
|Y\9 &=&
\sqrt{1-\Dx}|y\9|\xi_y\9+\sqrt{\Dx}|x\9|\zeta_y\9,
\label{XYeqs} \end{eqnarray}
and
\begin{eqnarray} |U\9 &=&
\sqrt{1-\Du}|u\9|\xi_u\9+\sqrt{\Du}|v\9|\zeta_u\9, \nonumber
\\ 
|V\9 &=&
\sqrt{1-\Du}|v\9|\xi_v\9+\sqrt{\Du}|u\9|\zeta_v\9,
\label{UVeqs} \end{eqnarray}
where each pair $|\xi_i\9$ and $|\zeta_i\9$ are normalized vectors
that are orthogonal to each other:
$\6\xi_x|\zeta_x\9=\6\xi_y|\zeta_y\9=\6\xi_u|\zeta_u\9=
\6\xi_v|\zeta_v\9=0$.

The remaining relations between the $|\xi_i\9$ and
$|\zeta_j\9$ cannot be chosen arbitrarily.  For instance, the 
orthogonality of $|X\9$ and $|Y\9$ requires that

\beq \6\xi_x|\zeta_y\9+\6\zeta_x|\xi_y\9=0. \label{Mingus} \eeq
Moreover, Eqs.~(\ref{conj}) and (\ref{CONJ}) imply that
\begin{eqnarray}
2\sqrt{1-\Du}|\xi_u\9 &=& 
\sqrt{1-\Dx}\,\Bigl(|\xi_x\9 + |\xi_y\9\Bigr) +
\sqrt{\Dx}\,\Bigl(|\zeta_x\9 + |\zeta_y\9\Bigr)  \nonumber\\
2\sqrt{\Du}|\zeta_u\9 &=&
\sqrt{1-\Dx}\,\Bigl(|\xi_x\9 - |\xi_y\9\Bigr) +
\sqrt{\Dx}\,\Bigl(|\zeta_y\9 - |\zeta_x\9\Bigr)\;,  
\label{YummaYumma} \end{eqnarray}
and similar relations for $|\xi_v\9$ and $|\zeta_v\9$. These in turn,
through $\6\xi_u|\zeta_u\9=\6\xi_v|\zeta_v\9=0$, lead to

\beq \mbox{Re}\, (\6\xi_x|\zeta_y\9-\6\zeta_x|\xi_y\9)=0,
\label{Getz} \eeq
and

\beq (1-\Dx)\,\mbox{Im}\,(\6\xi_y|\xi_x\9)+
\Dx\,\mbox{Im}\,(\6\zeta_x|\zeta_y\9)=0. \label{RubyBraff} \eeq
These requirements still leave us considerable freedom in the choice
of Eve's interaction with Alice and Bob's qubit.

Since Alice's input states only involve real coefficients, it is 
plausible that complex numbers are not necessary for describing 
Eve's optimal probe. We thus assume that 
all inner products between the various $|\xi_i\9$ and
$|\zeta_j\9$ are real numbers.  Then Eqs.~(\ref{Mingus}) and 
(\ref{Getz}), when combined, indicate that
$\6\xi_x|\zeta_y\9=\6\zeta_x|\xi_y\9=0$.

A particular choice for Eve's interaction that is adequate for our
needs can now be specified.  Recall that Eve's probe never need have
more than a four-dimensional Hilbert space.  That is to say, Eve's
probe may be taken to be two qubits.  It is therefore convenient to
introduce the same bases for each of Eve's qubits that we introduced
for Alice's qubit, namely $xy$ and $uv$.  In terms of these basis
vectors, we may further construct two standard (maximally) entangled
bases for the two qubits: a Bell basis \cite{Braunstein} with respect
to $xy$
\begin{eqnarray}
|\Phi^\pm_{xy}\9 &=& (|x\9|x\9\pm|y\9|y\9)/\sqrt{2} \nonumber\\
|\Psi^\pm_{xy}\9 &=& (|x\9|y\9\pm|y\9|x\9)/\sqrt{2},
\label{BillyKid} \end{eqnarray}
and similarly a Bell basis with respect to $uv$ consisting
of $|\Phi^\pm_{uv}\9$ and $|\Psi^\pm_{uv}\9$.

In terms of the Bell basis vectors for Eve's probe, we may choose the
interaction in such a way that
\begin{eqnarray}
|\xi_x\9&=&\sqrt{1-\Du}\,|\Phi^+_{xy}\9+\sqrt{\Du}\,|\Phi^-_{xy}\9,
\nonumber\\
|\xi_y\9&=&\sqrt{1-\Du}\,|\Phi^+_{xy}\9-\sqrt{\Du}\,|\Phi^-_{xy}\9,
\nonumber\\
|\zeta_x\9&=&\sqrt{1-\Du}\,|\Psi^+_{xy}\9-\sqrt{\Du}\,|\Psi^-_{xy}\9,
\nonumber\\
|\zeta_y\9&=&\sqrt{1-\Du}\,|\Psi^+_{xy}\9+\sqrt{\Du}\,|\Psi^-_{xy}\9.
\label{Reqs}
\end{eqnarray}
With respect to the conjugate inputs, the interaction takes a similar
form:
\begin{eqnarray}
|\xi_u\9&=&\sqrt{1-\Dx}\,|\Phi^+_{uv}\9+\sqrt{\Dx}\,|\Phi^-_{uv}\9,
\nonumber\\
|\xi_v\9&=&\sqrt{1-\Dx}\,|\Phi^+_{uv}\9-\sqrt{\Dx}\,|\Phi^-_{uv}\9,
\nonumber\\
|\zeta_u\9&=&\sqrt{1-\Dx}\,|\Psi^+_{uv}\9-\sqrt{\Dx}\,|\Psi^-_{uv}\9,
\nonumber\\
|\zeta_v\9&=&\sqrt{1-\Dx}\,|\Psi^+_{uv}\9+\sqrt{\Dx}\,|\Psi^-_{uv}\9.
\label{Seqs}
\end{eqnarray}
The second set of vectors is, of course, related to the first---as
it must be by unitarity---through relations such as in 
Eq.~(\ref{YummaYumma}).  Note that
neither collection of relative states is orthonormal.  Hence the set
of density operators available to Eve after the probe's
interaction---i.e., the set of quantum states from which she gains 
information about Alice's signal---is a noncommuting set.

To see that this interaction is optimal for Eve, we need only find
optimal POVMs $\{E_\1\}$ and $\{F_\1\}$---one for  each basis $xy$
and $uv$---to use under these  assumptions.  Then the optimality of
the whole procedure can be  checked either by testing the validity of
Eqs.~(\ref{Beebub})--(\ref{Condxy}), or simply by checking directly
that the bound is attained.  We opt for the former of these here. In
Sect.~IV, we shall use a direct check for a different set of
$|\xi_i\9$ and $|\zeta_i\9$.

Suppose Alice announces that a signal from the $xy$ basis was sent to
Bob. Then a natural choice for the observable Eve should measure is
the one that minimizes her error in guessing Alice's signal, i.e.,
the one that maximizes $G$ in Eq.~(\ref{G}). The corresponding basis
is well known~\cite{Helstrom76,Fuchs96}: it simply is the one that
diagonalizes the Hermitian operator

\beq \Gamma_{xy}=\rho_x-\rho_y, \eeq
where

\beq
\rho_x={\rm Tr}_{\rm Alice}(|X\9\6X|)=(1-
\Dx)|\xi_x\9\6\xi_x| + \Dx |\zeta_x\9\6\zeta_x|,
\eeq 
and likewise for $\rho_y$. The corresponding 
eigenprojectors of $\Gamma_{xy}$ are then given by

\beq
E\2 = |E\2\9\6E\2|,
\eeq
where
\beq
|E_0\9=|x\9|x\9, \qquad |E_1\9 = |y\9|x\9,\qquad
|E_2\9 = |x\9|y\9,\qquad |E_3\9 = |y\9|y\9.
\label{Eeqs} \eeq
Arbitrary values, 0, \ldots\, 3, have been assigned here to the label
$\1$.

Similarly, we make the analogous guess for Eve's measurement
in the case that Alice reveals the $uv$ basis; namely, we use the
eigenprojectors

\beq
  F\2 = |F\2\9\6F\2|
\eeq
of the operator

\beq
\Gamma_{uv}=\rho_u-\rho_v,
\eeq
where the density operators $\rho_u$ and $\rho_v$ are partial traces
of $|U\9$ and $|V\9$, respectively.  Again, it is easily verified 
that the appropriate eigenvectors are:
\beq
|F_0\9 = |u\9|u\9, \qquad |F_1\9 = |v\9|u\9, \qquad 
|F_2\9 = |u\9|v\9 \qquad |F_3\9 = |v\9|v\9.
\label{Feqs} \eeq

It should be noted that the measurement optimal for minimizing the 
error in a guess of the state's identity is generally {\it not\/} the 
same as the measurement for maximizing the mutual information 
about the state~\cite{Fuchs94}.  Thus there is no automatic 
guarantee that, even with the optimal interaction for Eve's probe, 
the measurements listed above will be adequate for achieving the
maximum possible mutual information.  Nevertheless for the case at 
hand, as will be seen shortly, circumstances have worked out in our 
favor.

With all the pieces in place, checking the optimality of the 
interaction given by Eqs.~(\ref{XYeqs}) and (\ref{Reqs}) and the
measurement given by Eq.~(\ref{Eeqs}), is just a question of checking 
that Eqs.~(\ref{Beebub})--(\ref{Condxy}) are satisfied.

We start by examining the vectors defined in Eq.~(\ref{UVdefs})
using the projectors onto the vectors of Eqs.~(\ref{Eeqs}). Note 
that in this case $\sqrt{E\2}=E\2$ is a matrix of rank 1.
Therefore, $B_u\0E\2$ projects onto a one-dimensional subspace of the
qubit-probe Hilbert space, so that
$|V_{\1u}\9$ and $|U_{\1u}\9$ are parallel.  Likewise
$|V_{\1v}\9$ and $|U_{\1v}\9$ are parallel. 
Working out the scaling factors between the parallel vectors is a
matter of applying the projectors to the
expressions in Eq.~(\ref{UVeqs}).  For example,
\begin{eqnarray}
  |U_{1u}\9 \,=\, B_u\0 E_1 |U\9 & = &
   \sqrt{1-\Du}\6E_1 | \xi_u\9 |u\9 |E_1\9, \nonumber \\
  & = & \sqrt{1-\Du} \sqrt{\Dx} |u\9 |E_1\9 / \sqrt{2},
\end{eqnarray}
and $|V_{1u}\9$ is given by the same expression except that
$\sqrt{1-\Du}$ is replaced by $\sqrt{\Du}$.  Hence 
Eq.~(\ref{Beebub}) is satisfied for $\1=1$ with $\epsilon_1=+1$.
One can work out the other cases in the same way, and show that

\beq
 \epsilon_0=+1,\qquad  \epsilon_1=+1,\qquad 
 \epsilon_2=-1,\qquad  \epsilon_3=-1.
\label{PresidentialDebates}
\eeq
Consequently, the measurement corresponding to Eq.~(\ref{Eeqs})
provides a mutual information $I_{xy}$ given by the right side
of Eq.~(\ref{ID}).  It is similarly straightforward to verify 
Eqs.~(\ref{HipHop}) and (\ref{Condxy}) by applying projectors
of the type $B_x \0 F\2$ and $B_y \0 F\2$ to the expressions in
Eq.~(\ref{XYeqs}), to form the quantities defined in
Eq.~(\ref{XYdefs}).

Hence there exists a definite choice of qubit-probe interaction,
namely Eqs.~(\ref{XYeqs})--(\ref{YummaYumma}), which, together with
two distinct measurement strategies, based upon Eqs.~(\ref{Eeqs}) and
(\ref{Feqs}) according to the basis announced by Alice, allows Eve to
saturate the bounds in Eqs.~(\ref{ID}) and (\ref{IDb})
simultaneously, for arbitrary choices of $\Du$ and $\Dx$.

As a final point, it is intriguing to note the following.  If Eve's 
concern were only to guess the state Alice prepared---and {\it not\/}
maximize her mutual information---then, clearly, it is enough for her
to bin the outcomes of her measurement two by two.  That is to say,
if Alice sends a signal in the $xy$ basis, then Eve upon receiving 
either outcome $E_0$ or $E_1$ should guess that the state $|x\9$
was sent; upon receiving either $E_2$ or $E_3$, she should guess that
$|y\9$ was sent.  These choices will minimize her probability of 
making an incorrect guess.  Similarly, she should guess $|u\9$ when 
she finds either $F_0$ or $F_1$ and  $|v\9$ when she finds either
$F_2$ or $F_3$.  Interestingly, Eq.~(\ref{PresidentialDebates}) along
with Eqs.~(\ref{Beebub}) and (\ref{Conduv}) (and similarly for the
conjugate basis) reveals that such a binned measurement is also
sufficient for maximizing Eve's mutual information.  Moreover, this
fact has another remarkable consequence: regardless of which basis
Alice used, after Eve's interaction, she can completely ignore the
first qubit of her probe.  All the accessible information about
Alice's signal is contained in the second qubit.  Thus, while two
qubits in Eve's probe are required for producing a minimal
disturbance interaction with Alice's qubit, only one qubit plays a
role in the final information-gathering process.  Also see the
discussion in the following paper~\cite{GN}.\bigskip
\vspace{5cm}

\begin{center}{\bf IV. OPTIMAL EAVESDROPPING STRATEGY}
\end{center}\medskip

We are finally in a position to describe the eavesdropping strategy 
that is most relevant to quantum cryptography with the BB84 protocol.
Namely, we should like to know Eve's best {\it average\/} mutual
information for a  fixed {\it average\/} disturbance across the two
bases $xy$ and $uv$. This is given by combining the two results of 
Eqs.~(\ref{ID}) and (\ref{IDb}).  Fixing the average disturbance to
be

\beq
D=\half\,(D_{xy}+D_{uv}),
\eeq
and defining

\beq G=\half\,(G_{xy}+G_{uv})\qquad{\rm and}\qquad
 I=\half\,(I_{xy}+I_{uv})
\label{GDav}
\eeq
for the average information gain and mutual information,
respectively, we can again use the concavity of the functions
$[x(1-x)]^{1/2}$ and $\phi\Bigl[2\sqrt{x(1-x)}\,\Bigr]$ to obtain

\beq G\leq2\,[D\,(1-D)]^{1/2},
\label{GD}
\eeq
and

\beq
I \leq \half\,\phi\Bigl[ 2\sqrt{D\,(1-D)}\,\Bigr].
\label{IDc}
\eeq
                 
Equality can be achieved in either of these bounds only if

\beq
D_{xy}=D_{uv}=D.
\eeq
The result is plotted in Fig.~3. As intuitively expected, the average
error is the same in both channels. If it were not so, different
error rates for $xy$ and $uv$ signals would be a telltale indication
that a clumsy eavesdropper is tampering with the communication line.

The derivation of Eq.~(\ref{IDc}) as given above may seem long and 
arduous. This is due to the
generality of the previous sections:  Section~III encompasses 
strategies that produce asymmetric disturbances in the two conjugate 
bases and the bounding argument of Section~II can, with slight
modification, be generalized to nonconjugate bases and unequal prior 
probabilities for those bases.  To more firmly place the physics of 
the optimal eavesdropping strategy in Eq.~(\ref{IDc}) within 
context, we now sketch an alternate derivation based on a
symmetrization argument.

The starting point of the new argument is to notice that for any
eavesdropping procedure Eve chooses to use, there exists a
symmetrized strategy leading to the same average information $I$, and
same {\it or lesser\/} average disturbance $D$. For each one of the
signals sent by Alice, the mixed states Bob receives can be made to
be of the form

\begin{equation}
\rho_{\rm Bob}=(1-2D)\,\rho_{\rm Alice}+D\,{\bf 1}, \label{damneq}
\end{equation}
as if Alice's signals were merely diluted by mixing them with a
random component. A formal proof of this result is given in
Appendix B.

Therefore with no loss of generality we can obtain Eve's ultimate
bound on information versus disturbance by studying symmetric 
strategies.  Note, however, that this may come at the cost of 
adding extra degrees of freedom to Eve's setup:  without these, we
would not be able to enact the required random orientation.  For 
instance, if Eve's probe were restricted to consist of a single
qubit, as in Ref.~\cite{GH}, 
there would be no way to carry out this symmetrization.  However, 
by making no a priori restrictions on Eve's probe, symmetrized 
strategies can always be covered within our formal framework.  In 
particular, there must exist an optimal strategy on Eve's part that 
gives Eqs.~(\ref{XYeqs}) and (\ref{UVeqs}) with $\Dx = \Du = D$.

Again, on physical grounds, it is plausible that the Schmidt states
in Eve's probe are real (not complex) superpositions with respect to
some basis, as in Eqs.~(\ref{XYeqs}) and (\ref{UVeqs}).  (Actually it
can be checked that no new result is obtained by introducing complex
coefficients.  For the sake of brevity, however, we consider only
real coefficients in the following.) Then running through the same
argument as presented between Eqs.~(\ref{Mingus}) and (\ref{RubyBraff})
and in the following paragraph, we have

\beq
\6\xi_x|\zeta_x\9=\6\xi_y|\zeta_y\9=
\6\xi_x|\zeta_y\9=\6\xi_y|\zeta_x\9=0.
\eeq

These requirements are enough to ensure that the set of relevant
$|\xi_x\9$, $|\zeta_x\9$, $|\xi_y\9$, and $|\zeta_y\9$ can all be
parameterized by two real numbers. There are now many possibilities
open. Instead of (\ref{Reqs}), we may try a solution that looks
simpler,
such as
\begin{eqnarray}
|\xi_x\9 &=& |x\9|x\9, \nonumber\\
|\zeta_x\9 &=& |x\9|y\9, \nonumber\\
|\xi_y\9 &=& (\cos\alpha|x\9+\sin\alpha|y\9)|x\9, \nonumber\\
|\zeta_y\9 &=& (\cos\beta|x\9+\sin\beta|y\9)|y\9.
\label{TiredBack}
\end{eqnarray}
It then follows from $\6\zeta_u|\zeta_u\9=1$ that

\beq
D=\frac{1-\cos\alpha}{2-\cos\alpha+\cos\beta}\;.
\eeq

Let us consider the case where Alice announces that the $xy$
basis has been used.  Then the two density operators that Eve must
distinguish are
\begin{eqnarray}
\rho_x &=& (1-D)|\xi_x\9\6\xi_x|+D|\zeta_x\9\6\zeta_x|,
\nonumber\\
\rho_y &=& (1-D)|\xi_y\9\6\xi_y|+D|\zeta_y\9\6\zeta_y|.
\end{eqnarray}
The optimal
information gathering measurement for these two states proceeds as
follows: Eve first performs a preliminary step of distinguishing the
vectors---rather than the density operators---by measuring the second
qubit, because the set of $|\xi_i\9$ are orthogonal to the set of
$|\zeta_j\9$.  The set of $|\xi_i\9$ will occur with probability
$(1-D)$; the set of $|\zeta_j\9$ will occur with probability $D$.
Thereafter, distinguishing the density operators $\rho_x$ and
$\rho_y$ becomes a question of distinguishing the (equiprobable) pure
states in the appropriate set.  The optimal information gathering 
measurement in either case is defined by the basis that straddles the 
two nonorthogonal vectors that must be distinguished \cite{Levitin}.
In the two cases, this leads to an information gain on
Eve's part given by \cite{Levitin,Fuchs94}
\begin{eqnarray}
I_\xi &=& \half(1+\sin\alpha)\ln(1+\sin\alpha)+
\half(1-\sin\alpha)\ln(1-\sin\alpha),
\nonumber\\
I_\zeta &=& \half(1+\sin\beta)\ln(1+\sin\beta)+
\half(1-\sin\beta)\ln(1-\sin\beta).
\end{eqnarray}
On average, Eve's information gain is given by

\beq
I=(1-D)\,I_\xi + D\,I_\zeta.
\eeq

Eve's optimal strategy is obtained with the values of $\alpha$ and
$\beta$ that maximize $I$ when $D$ is fixed.  One can readily check
that this occurs when $\alpha=\beta$, with
$\sin\alpha=2\sqrt{D(1-D)}$.
By symmetry, the same result holds when Alice reveals that the $uv$ 
basis has been sent (though the detailed protocol for measuring the
two qubits is slightly different in that case). We again find
Eq.~(\ref{IDc}) as the optimal information--disturbance tradeoff.

For small values of $D$, the bound given in Eq.~(\ref{IDc}) becomes
$I\leq2D$.  At the other extreme, the maximum value of $I$ is
$\ln 2$ (that is, one bit): Eve can achieve this result simply by
keeping Alice's qubit for herself, and sending to Bob a dummy qubit 
in a random state.
She then has all the information, and Bob gets a 50\% error rate.
This state of affairs should be contrasted to optimal eavesdropping
on the quantum cryptographic protocol B92 of Bennett~\cite{B92},
that uses only two nonorthogonal quantum states. There one finds,
for small values of $D$, that $I\propto\sqrt{D}$ \cite{FP}.  This
suggests that the BB84 protocol is inherently more secure against
eavesdropping than the B92 scheme: for a given disturbance, Eve
obtains more information about the identity of Alice and Bob's bit
in B92 than in BB84.

To this point, we have hardly discussed what Alice and Bob can do
with the knowledge of Eq.~(\ref{IDc}) and Eve's optimal strategy
(given our restrictions to the problem).  Generally, the users of
the BB84 protocol will not have a noiseless communication channel
available for their use. If Alice and Bob use a noisy channel, the
only truly safe way for them to proceed is to assume that all the
noise is due to some Eve using an optimal eavesdropping scheme.
Then, if this Eve has not been too invasive, Alice and Bob may still
be able to recover a safe cryptographic key by methods of privacy 
amplification.

As discussed in Refs.~\cite{EHPP,Maurer}, a good indicator of Alice
and Bob's capability of recovering a safe cryptographic key in the
face of Eve's presence can be formulated in terms of various mutual
informations. In particular, one must compare the mutual information
$I_{AB}$ between 
Alice and Bob (after Eve's eavesdropping) to the mutual informations 
$I_{AE}$ and $I_{EB}$ between Alice and Eve and between Eve and Bob, 
respectively.  If the natural noise in the channel is such that
$I_{AB}\le\min\{I_{AE},I_{EB}\}$, for any potential eavesdropper, 
then Alice and Bob should consider the channel inappropriate for
quantum cryptographic key generation.  They should either move to 
another channel or give up their quest.

Note that for the optimal scheme derived here $I_{AE}=I_{EB}$ and
both are given by the right hand side of Eq.~(\ref{IDc}).  On the
other hand, as far as Alice and Bob are concerned, Eve's action has
merely produced a binary symmetric channel between them, with a
data-flipping rate $D$.  Therefore~\cite{Cover}
\beq I_{AB} = \ln2 + D\ln D + (1-D)\ln(1-D)
= \half\,\phi(1-2D). \eeq
Comparing this expression to Eq.~(\ref{IDc}), we can find the
threshold noise level for a potentially safe channel; namely, it
occurs when

\beq
|1-2D|=2\sqrt{D(1-D)}.
\eeq
That is to say, when

\beq
D\ge\half-\mbox{$1\over4$}\sqrt{2}\approx 0.146447,
\label{TooMuchCoffee}
\eeq
the channel should be considered too risky for safe key generation.

Finally, let us discuss an intriguing connection between optimal 
eavesdropping and the violation of Bell inequalities.  A slight 
modification of the BB84 protocol can be built upon Alice and Bob
sharing an entangled pair of qubits (such as the singlet state
$|\Psi^-\9$) rather than Alice physically sending a qubit to
Bob~\cite{BBM}.  Alice and Bob simply randomly perform measurements
in the $xy$ and $uv$ bases, and announce their measurement---though
not their result---to each other.  Whenever their measurement bases
differ, they discard the bit; whenever the bases are 
the same, they know that they should have opposite bits if there were
no eavesdropping or noise on the channel.  An eavesdropper in this
scenario might be imagined to interact with one qubit of the EPR pair
in an attempt to gather information about Alice and Bob's final key.

Ekert~\cite{Ekert91}, in a related scheme, pointed out that an 
appropriate test for eavesdropping might be a check on whether
the Bell inequalities are violated. This can 
be enacted in our scenario by allowing Bob to rotate his measuring 
apparatus by 22.5 degrees.  Then Alice and Bob will be in position
for testing the standard Clauser-Horne-Shimony-Holt (CHSH)
inequality~\cite{CHSH}.  
The correlation signature $S$ in that inequality cannot exceed
2 for theories based on local hidden variables.  However, in the
modified BB84 protocol just discussed, $S$ can reach $2\sqrt{2}$ when 
there is no eavesdropping involved.  The effect on $S$ of
our optimal eavesdropping strategy is equivalent to the one caused
by a data-flipping error with probability $D$ in one of the
detectors~\cite{Braunstein2}:

\beq
S=2\sqrt{2}\,(1-2D).
\eeq
It is noteworthy that the CHSH inequality ceases to be violated, 
i.e., $S\le2$,  just when $D$ satisfies Eq.~(\ref{TooMuchCoffee}).
This confirms the conjecture of Gisin and Huttner \cite{GH} and to
some extent vindicates the idea of Ekert.  We believe this connection
between privacy amplification requirements and Bell inequalities 
may have fundamental implications in quantum information theory and
is worthy of further investigation.
\bigskip
\vspace{5cm}

\begin{center}{\bf ACKNOWLEDGMENTS}\end{center}\medskip

We thank the Institute for Theoretical Physics at the University of
California at Santa Barbara (under NSF Grant No.\ PHY94-07194) for
their hospitality while this work was being executed.  CAF thanks
G.~Brassard for many discussions and acknowledges the receipt of the
Lee A. DuBridge Fellowship and the support of DARPA through the
Quantum Information and Computing (QUIC) Institute administered by
ARO.  NG acknowledges financial support by the Swiss National Science
Foundation and the European TMR Network on the Physics of Quantum
Information.  RBG and CSN acknowledge the support of NSF and ARPA
through grant CCR-9633102.\bigskip

\begin{center}{\bf APPENDIX A. PROOF OF CONCAVITY}\end{center}
\medskip

Consider the function

\beq \phi(z)=(1+z)\ln(1+z)+(1-z)\ln(1-z).\eeq
We have 

\beq \phi'(z)=\ln[(1+z)/(1-z)], \eeq
and

\beq \phi''(z)=2/(1-z^2). \eeq
Now let

\beq z(x)=2\,[x\,(1-x)]^{1/2}, \eeq
whence

\beq z'(x)=(1-2x)/[x\,(1-x)]^{1/2}, \eeq
and

\beq z''(x)=-\half\,[x\,(1-x)]^{-3/2}=-4/z^3.\eeq
We have

\beq \frac{d\phi}{dx}=\frac{d\phi}{dz}\,\frac{dz}{dx}, \eeq
whence

\beq \frac{d^2\phi}{dx^2}=\frac{d\phi}{dz}\,\frac{d^2z}{dx^2}+
 \frac{d^2\phi}{dz^2}\,\left(\frac{dz}{dx}\right)^2. \eeq
Combining all these equations together, we obtain

\beq \frac{d^2\phi}{dx^2}=\frac{4}{z^3}\,\left(2z-\ln\frac{1+z}{1-z}
  \right).
\eeq
Recall that $0<z<1$. The parenthesis on the right hand side of
(\theequation) vanishes for $z=0$, and its derivative is
$2-2/(1-z^2)$, which is always negative. Therefore
$(d^2\phi/dx^2)<0$, and it follows that the function $\phi[z(x)]$ is
concave.\bigskip

\begin{center}{\bf APPENDIX B. SYMMETRIZED EAVESDROPPING}\end{center}
\medskip

The purpose of this Appendix is to prove Eq.~(\ref{damneq}). Consider
the representation of Alice's four states on a Poincar\'e sphere.
They lie on the equatorial plane, at the ends of two perpendicular
diameters. The states that Bob receives are also represented by four
points. The latter are located {\it inside\/} the sphere, since these
are mixed states.

Eve proceeds as follows: before eavesdropping, she randomly rotates
Alice's signal by 0, 45, 90, or 135 degrees in the plane of Fig. 1
(that is, she rotates the Poincar\'e sphere by 0, 90, 180, or 270
degrees around its polar axis). After the eavesdropping
interaction, she rotates the signal back, and then sends it to Bob.
This causes no change to the {\it average\/} amount of information
she gathers, but equalizes the disturbances to Alice's four states.
By virtue of this symmetrization, the set of Bob's states is now
invariant under rotations of the Poincar\'e sphere by 90, 180, and
270 degrees. Therefore, the four points representing these states
form a square, lying in a plane parallel to the equatorial plane. 
If the sides of that square are not parallel to those of the square
formed by Alice's states, they can be made parallel by a further
rotation around the polar axis. This does not change Eve's $I$, but
this reduces Bob's $D$, thus improving the eavesdropping method.

Moreover, the four points that represent $\rho_{\rm Bob}$ can be made
to lie on the equatorial plane itself, not on a parallel plane above
or below it. If they are not on the equatorial plane, this means that
the eavesdropping interaction produces a circularly polarized
component in the outgoing state (recall that the poles of the
Poincar\'e sphere represent pure circular polarizations). This is
indeed possible if the unitary interaction of the probe involves
complex coefficients. In that case, Eve ought to have two available
probes, whose interactions are described by complex conjugate unitary
matrices. The second probe yields Bob's states on the other side of
the equatorial plane. By randomly choosing one of the two probes, Eve
can bring Bob's states back to the equatorial plane (where Alice's
states are).  This changes neither $I$ nor $D$.

This argument proves that the result stated in Eq.~(\ref{damneq}) 
can indeed be achieved by symmetrizing any eavesdropping
strategy.  In particular, there must also be an optimal
strategy giving rise to Eq.~(\ref{damneq}).

\vfill

\noindent FIG. 1. \ The orthogonal bases $xy$ and $uv$,
that satisfy Eq.~(\ref{conj}), are called {\it conjugate\/} to each
other.\bigskip

\noindent FIG. 2. \ Information vs.~disturbance for various
eavesdropping methods.\bigskip

\noindent FIG. 3. \ Eve's information gain $G$ and mutual
information $I$ (in bits) as functions of Bob's error rate $D$.
\end{document}